\documentclass[sigconf]{acmart}
\AtBeginDocument{%
  }

\usepackage{graphicx} 
\usepackage{amsmath}
\usepackage{enumitem}

\usepackage{booktabs}
\usepackage{multirow}
\usepackage{makecell}
\usepackage[table]{xcolor}
\usepackage{colortbl}
\usepackage[most]{tcolorbox} 
\usepackage{listings}        
\tcbuselibrary{breakable}
\usepackage{nicematrix}

\copyrightyear{2026}
\acmYear{2026}
\setcopyright{cc}
\setcctype{by-nc-nd}
\acmConference[KDD '26]{Proceedings of the 32nd ACM SIGKDD Conference on Knowledge Discovery and Data Mining V.2}{August 09--13, 2026}{Jeju Island, Republic of Korea}
\acmBooktitle{Proceedings of the 32nd ACM SIGKDD Conference on Knowledge Discovery and Data Mining V.2 (KDD '26), August 09--13, 2026, Jeju Island, Republic of Korea}
\acmDOI{10.1145/3770855.3818407}
\acmISBN{979-8-4007-2259-2/2026/08}

\begin{document}

\title{Fighting Numerical Hallucinations via Data-centric Compilation for Online Financial QA}


\author{Hao Chen}
\authornote{Both authors contributed equally to this research.}
\affiliation{%
  \institution{Shenzhen Technology University}
  \country{}
  }
  \affiliation{
  \institution{FiT, Tencent}
  \city{Shenzhen}
  \country{China}
}

\author{Xing Tang}
\authornotemark[1]
\affiliation{%
  \institution{Shenzhen Technology University}
  \city{Shenzhen}
  \country{China}
}

\author{Qirui Liu}
\authornote{Work done when he was an intern at FiT, Tencent.}
\affiliation{
  \institution{South China University of Technology}
  \city{Guangzhou}
  \country{China}
}

\author{Weijie Shi}
\affiliation{
  \institution{The Hong Kong University of Science and Technology}
  \city{Hong Kong SAR}
  \country{China}
}

\author{Shiwei Li}
\affiliation{%
  \institution{Huazhong University of Science and Technology}
  \city{Wuhan}
  \country{China}
}

\author{Fuyuan Lyu}
\affiliation{
  \institution{McGill University}
  \city{Montreal}
  \country{Canada}
}

\author{Weihong Luo}
\affiliation{%
  \institution{FiT, Tencent}
  \city{Shenzhen}
  \country{China}
}

\author{Xiku Du}
\affiliation{%
  \institution{FiT, Tencent}
  \city{Shenzhen}
  \country{China}
}

\author{Xiuqiang He}
\authornote{Corresponding author.}
\affiliation{
  \institution{Shenzhen Technology University}
  \city{Shenzhen}
  \country{China}
}
\renewcommand{\shortauthors}{Hao Chen et al.}

\begin{abstract}

Large Language Models (LLMs) have significantly advanced online data services, particularly in the domain of financial question answering (FinQA). However, such systems remain susceptible to numerical reasoning hallucinations, which critically undermine reliability in high-stakes financial applications. Although retrieval-augmented generation (RAG) has been widely adopted to ground responses in external knowledge, it introduces three persistent challenges: noise sensitivity, calculation fragility, and an auditability crisis. Existing model-centric approaches, which primarily focus on optimizing either the retriever or generator in isolation, still struggle to address these issues in an integrated manner.
In this work, we pioneer a data-centric paradigm and propose a novel framework, the Data-centric Reasoning Compiler (DCRC). The framework operates through three cohesive phases: (1) adversarial data construction, which synthesizes training examples with controlled noise to teach robustness; (2) multi-stage training that cultivates a Data-centric Structuring Agent (DSA) capable of explicit evidence auditing and program synthesis; and (3) a compile-and-execute inference process, where the DSA transforms user queries and retrieved documents into verifiable, executable reasoning programs. This data-driven framework ensures faithful numerical reasoning by design.
We conduct extensive experiments on established offline benchmarks and further validate our framework through deployment in a real-world online financial QA system. Notably, the framework has also been adopted in the financial QA of YuanBao\footnote{https://yuanbao.tencent.com/chat/}, one of the largest chat platforms in China.
\end{abstract}


\maketitle

\section{Introduction}

\begin{figure*}
    \centering
    \includegraphics[width=0.9\linewidth]{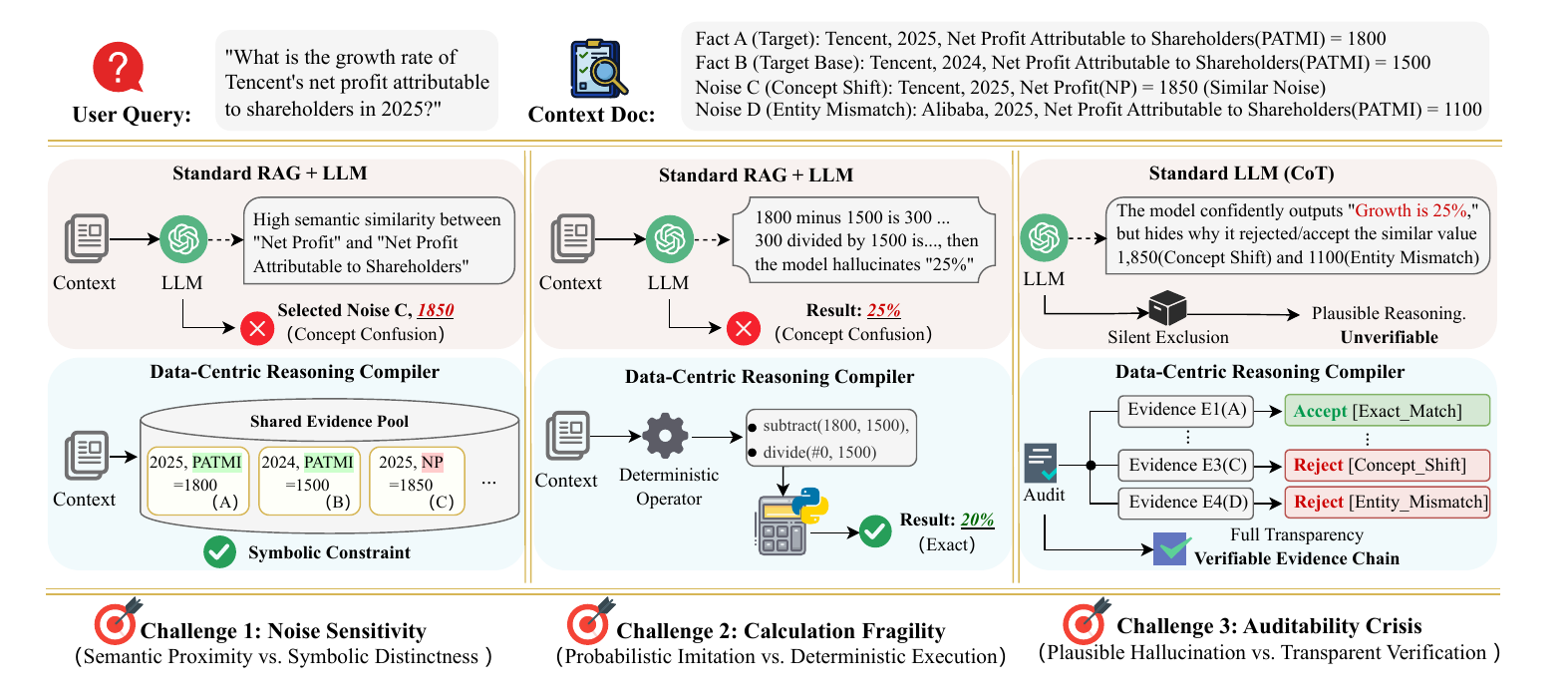}
    \caption{Illustrating the \textbf{DCRC} framework versus standard methods on a financial QA example. Standard approaches suffer from \emph{noise sensitivity} (confusing similar facts), \emph{calculation fragility} (stochastic generation), and \emph{auditability crisis} (opaque reasoning). DCRC overcomes these by constructing an adversarially-augmented evidence pool, synthesizing a deterministic operator program, and generating a fully verifiable audit log.}
    \label{fig:intro}
\end{figure*}

Large language models (LLMs)~\cite{gpt,qwen} have great potential to transform online data services. A prominent and impactful application is online financial question answering (FinQA), where LLMs enhance the interpretation and response to complex user queries. However, a core challenge in deploying such systems effectively is generating answers that are both \emph{numerically faithful} and \emph{explainable}~\cite{survey}, which requires robust multi-step reasoning. The recent advent of Large Reasoning Models (LRMs)~\cite{deepseek,qwen3,kimi,LRM}, which decompose complex problems into explicit reasoning chains,  provides a critical capability for addressing this challenge and building reliable online FinQA systems.

Nevertheless, the practical deployment of LRM-based FinQA systems is significantly hindered by the issue of \emph{numerical hallucinations}~\cite{hallucinations,deficiency}, which undermines the reliability of generated answers. To mitigate this, Retrieval-Augmented Generation (RAG)~\cite{rag} has emerged as a predominant technique, grounding model responses in external, up-to-date information. However, a critical limitation persists: even with access to retrieved evidence, LRMs may still produce \emph{unsupported reasoning steps} or present claims that \emph{contradict} the provided content~\cite{ragtruth}. This gap between retrieval and faithful reasoning directly impedes the development of truly robust and trustworthy FinQA systems. 
As shown in Figure~\ref{fig:intro}, a user queries the growth rate of Tencent's 2025 net profit attributable to shareholders. Standard RAG retrieves relevant facts but also introduces similar yet distinct noise; for example, the retrieved documents focus on net profit rather than net profit attributable to shareholders, which is semantically very similar. 
When relying on such noisy context, LLMs tend to incorrectly calculate the growth rate due to their inherent nature of probabilistic token prediction, which struggles to enforce deterministic arithmetic operations and faithfully align with exact numeric evidence. Even with chain-of-thought prompting~\cite{chain}, the model may silently omit relevant evidence without providing any verifiable reasoning to explain this exclusion. 

Most existing studies on numerical reasoning adopt a retriever-generator framework~\cite{apollo,encore,finder}, which decomposes the task into two stages: retrieval and generation. These methods focus on optimizing these two stages respectively. For example, fintuning the retriever~\cite{apollo}, optimizing the generator~\cite{encore}, and prompting strategies~\cite{finder}. These methods are still limited by the decoupled design. As a result, such approaches inherently fail to address the core challenges in a unified manner. Specifically, the retriever may retrieve noisy or irrelevant facts without explicit detecting potential negative samples, leading to \emph{Noise Sensitivity}; the generator relies on probabilistic next token generation rather than deterministic computation, a problem we term \emph{Calculation Fragility}; and the reasoning process lacks verifiable evidence chains, only leaving users with unverified answers that cause \emph{Auditability Crisis}.

In this work, we propose a paradigm shift from a model-centric to a \textbf{data-centric} approach, and introduce the \textbf{D}ata-\textbf{c}entric \textbf{R}easoning \textbf{C}ompiler (DCRC) framework, which trains a data-centric structuring agent (DSA) to enable a trustworthy “compile-and-execute” process. First, our framework begins by constructing a high-quality, adversarially augmented dataset through heterogeneous consensus mining and controlled noise injection, thereby teaching the DSA to directly discriminate and reject misleading facts, effectively addressing noise sensitivity. We then train the DSA with a multi-stage training. The structure-aware supervised fine-tuning (SFT) emphasizes critical auditing decisions, while reinforcement learning with verifiable multi-dimensional rewards ensures the generation of executable programs, thus eliminating calculation fragility. Finally, the trained DSA acts as a deterministic compiler at inference. DSA decomposes queries, audits evidence with explicit justifications, synthesizes structured programs, and executes them via a symbolic operator registry. The entire process yields a complete, verifiable audit log, providing full transparency and auditability. In summary, our main contributions are as follows:

\begin{itemize}[itemsep=0pt,parsep=0pt,leftmargin=*]
\item We first adopt a data-centric approach to trustworthy financial QA, which provides a practical and robust solution for online FinQA systems.

\item We introduce the Data-centric Reasoning Compiler, a novel framework comprising three cohesive phases: adversarial data construction, multi-stage training, and compile-and-execute inference. DCRC transforms standard LRMs into reliable reasoning engines that effectively combat numerical hallucinations through verifiable program synthesis.

\item Extensive experiments on both the offline dataset and the online scenario have been conducted to validate the superiority of our method.
\end{itemize}
\section{Related work}
In this section, we review two topics related to our work, including financial QA and numerical reasoning.

\subsection{Financial QA}

The rapid advancement of Large Language Models (LLMs) has substantially enhanced online question-answering (QA) systems, making them a core component of modern data services~\cite{webglm,explore,qasurvey,uqabench,data}. Financial QA, as a critical application domain, has particularly benefited from these developments, enabling users to access and analyze complex financial information in a timely manner~\cite{survey}. 
Research in this field can be broadly categorized into two directions. The first focuses on financial numerical QA, which involves multi-step calculations and the extraction of relevant information from diverse data sources (e.g., tables, text)~\cite{financereasoning,tat,finben}. This includes zero-shot techniques that leverage LLMs' inherent reasoning capabilities~\cite{zeroshot} to handle complex queries without task-specific training~\cite{zshot,finfinetuning}, as well as multi-agent approaches that employ critic models to reflect on reasoning steps and answers~\cite{multi-agent}. The second direction concerns financial QA over pure text, where systems such as WeaverBird~\cite{weaverbird}, a dialogue system tailored to the finance sector, leverage LLMs fine-tuned on large financial corpora to provide informed responses to user inquiries.
Our work focuses on advancing numerical reasoning over retrieved tabular and textual content, aiming to provide a robust and complementary enhancement to existing financial QA systems.

\subsection{Numerical Reasoning}

Numerical reasoning remains a persistent challenge for large language models~\cite{finmmr,exposing, exploring}, and its reliable execution is central to robust financial question answering~\cite{apollo,encore,finder,finqa}. To advance the field, several benchmarks have been developed. FinQA~\cite{finqa} and ConvFinQA~\cite{convfinqa} target long-form numerical reasoning over financial reports, while FinMMR~\cite{finmmr} is a bilingual, multimodal benchmark designed to evaluate numerical reasoning capabilities in a financial context.
Motivated by these benchmarks, a line of research has emerged within the prevalent retriever-generator framework. APOLLO~\cite{apollo} enhances the retriever via number-aware negative sampling to better discriminate key numerical facts. ENCORE~\cite{encore} derives reliable reasoning processes by decomposing answer formulas. FINDER~\cite{finder} employs a generative retriever for fact extraction and uses dynamic in-context example selection to improve Program-of-Thought prompting. Our proposed framework shifts to a data-centric paradigm and transforms standard LRMs into reliable reasoning
engines.
\section{Method}

\begin{figure*}
    \centering
    \includegraphics[width=1\linewidth]{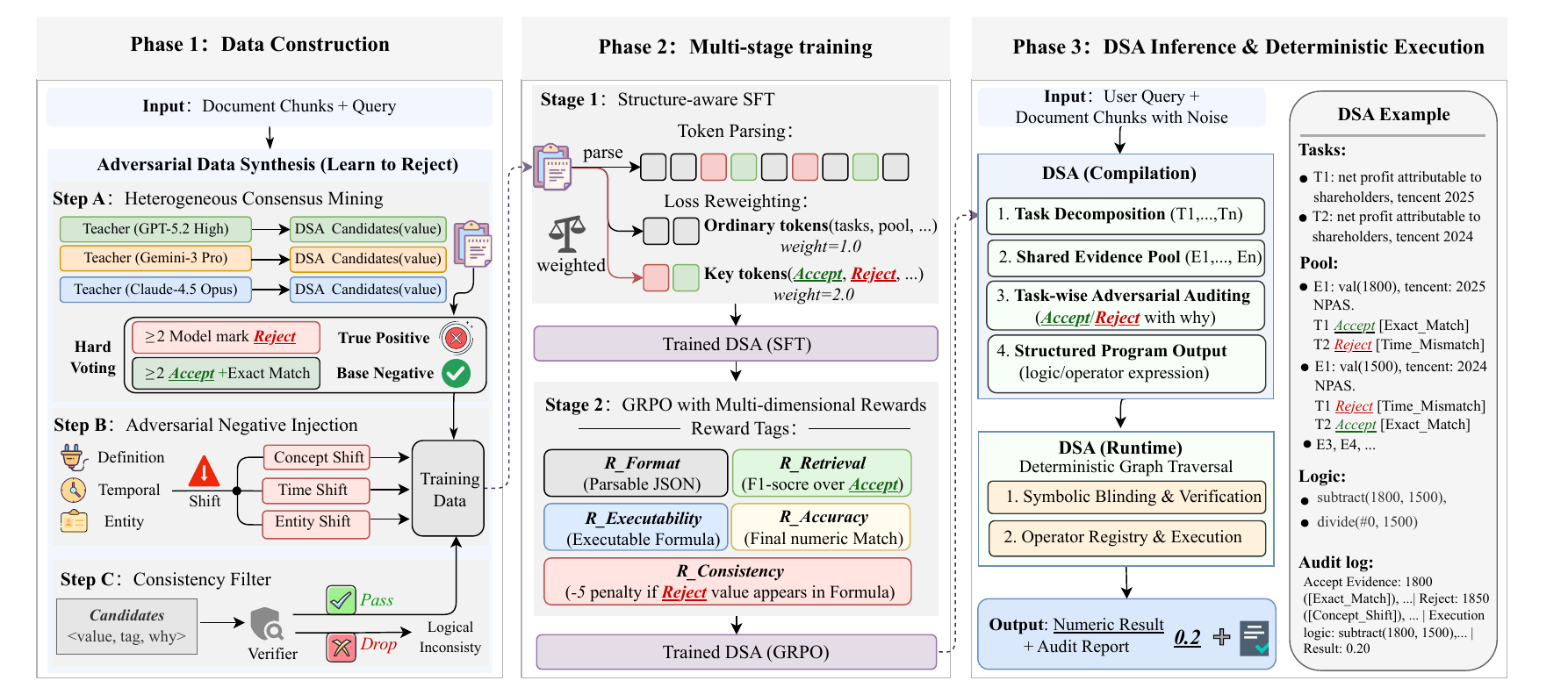}
    \caption{The overall architecture of the Data-centric Reasoning Compiler (DCRC) framework, consisting of three cohesive phases: (1) Adversarial Data Construction through heterogeneous consensus mining and hard negative injection; (2) Multi-Stage Training with structure-aware SFT and GRPO with multi-dimensional rewards; (3) Compile-and-Execute Inference where DSA generates verifiable intermediate representations executed by the deterministic CEA.}
    \label{fig:framework}
\end{figure*}

First, we provide a formal definition of our problem. Next, we elaborate on the proposed Data-centric Reasoning Compiler (DCRC) framework, which comprises three cohesive phases: adversarial data construction, multi-stage model training, and compile-and-execute inference. We will describe the design of each phase in detail.

\subsection{Problem Formulation}

Consider an online financial QA system powered by a Large Reasoning Model $M$. Given a user query $q$ and a set of retrieved document chunks $\mathbb{D}=\{d_1, d_2, \cdots, d_n\}$, the system is required to generate a numerically accurate and auditable answer $a$. Each document chunk $d_i$ can be represented as a tuple $d_i = (text_i, \mathcal{V}_i)$, where $text_i$ denotes the textual content and $\mathcal{V}_i = \{v_1, v_2, \cdots\}$ represents the set of candidate numerical values contained therein.

In the standard retrieval-augmented generation paradigm, the model directly generates the answer $a_g = M(q, \mathbb{D})$. However, this process is susceptible to numerical hallucinations due to noise sensitivity, calculation fragility, and auditability crisis.

Our goal is to train a Data-centric Structuring Agent (DSA) $\pi$ that compiles the query and documents into a verifiable intermediate representation $\mathcal{I}$, which is subsequently executed by a deterministic executor $\mathcal{E}$ to obtain the final answer:
\begin{equation}
a = \mathcal{E}(\mathcal{I}), \quad \mathcal{I} = \pi(q, \mathbb{D})
\end{equation}

The intermediate representation $\mathcal{I}$ adheres to our designed \textbf{DSA Schema}, which mandates: (1) sub-task decomposition $\mathcal{T} = \{T_1, T_2, \cdots\}$; (2) an evidence pool $\mathcal{P}$ with explicit auditing decisions; and (3) symbolic computation logic $\mathcal{L}$. This design ensures complete traceability of the reasoning process.

\subsection{Adversarial Data Construction}

In our DCRC framework, high-quality data is key. The data includes not only positive samples but also negative samples.
To teach the model to ``learn to reject'' misleading evidence, we propose a systematic adversarial data synthesis pipeline comprising three steps: heterogeneous consensus mining, adversarial negative injection, and consistency filter.

\subsubsection{Heterogeneous Consensus Mining}

To eliminate single-model bias and hallucinations, we construct a teacher committee of $K$ heterogeneous models $\{M_1, \ldots, M_K\}$ (e.g., GPT-5.2 High~\footnote{https://chatgpt.com/}, Gemini-3 Pro~\footnote{https://gemini.google.com/}, Claude-4.5 Opus~\footnote{https://claude.ai/}). For each query-document pair $(q, \mathbb{D})$, each model independently generates a structured DSA representation $\mathcal{I}_k = M_k(q, \mathbb{D})$. We then apply hard voting to determine golden labels: for each candidate value $v$,
\begin{equation}
\text{Consensus}(v) = \begin{cases}
\text{Accept}, & \text{if } \sum_{k=1}^{K} \mathbb{I}[\text{tag}_k(v) = \text{Accept}] \geq \lceil K/2 \rceil \\
\text{Reject}, & \text{otherwise}
\end{cases}
\end{equation}
where $\text{tag}_k(v)$ denotes the $k$-th model's decision. Note that only when the majority agrees($\geq \lceil K/2 \rceil$), and the extracted values are identical, is the evidence included in the positive set $\mathbb{B}^+$.

\subsubsection{Hard Negative Injection}

Instead of merely adding positive examples, we systematically inject high-difficulty negative samples that mimic real-world error patterns. This forces the model to learn not only what information to use but also what to reject, laying the foundation for verifiable reasoning. Based on positive samples, we generate high-difficulty adversarial negative samples through controlled perturbation, covering three core error patterns. Concept Shift identifies values in the document that are semantically similar to but definitionally different from the target metric; for instance, when the target is ``net profit attributable to shareholders,'' we construct negative samples using ``operating profit'' values, annotated with reason \texttt{[Concept\_Shift]}. Temporal Mismatch maintains the metric type but substitutes values from different time periods; for example, if the query requires ``2023'' data, we inject values from ``2022'' as distractors, annotated with reason \texttt{[Time\_Mismatch]}. Entity Confusion introduces values from different entities (e.g., subsidiary vs. group) as negative samples, annotated with reason \texttt{[Entity\_Mismatch]}.

Let the original positive sample evidence be $e^+ = (v, \text{src}, \text{ACCEPT})$, where $v$ denotes the extracted numerical value and $\text{src}$ represents the source text snippet from the original document where the value was found. The perturbation function $\phi_{\tau}$ generates negative samples according to perturbation type $\tau \in \{\text{concept}, \text{temporal}, \text{entity}\}$:
\begin{equation}
e^- = \phi_{\tau}(e^+, \mathbb{D}) = (v', \text{src}', \text{Reject}, \text{reason}_\tau)
\end{equation}
where $v'$ is the perturbed distractor value, $\text{src}'$ is the corresponding source text, and $\text{reason}_\tau$ is a predefined structured error code.

\subsubsection{Consistency Filtering}

To ensure the logical self-consistency of synthesized data, we introduce a verification model $M_v$ for final cleansing. For each sample, we check the semantic consistency between its tag and reason:
\begin{equation}
\text{Valid}(e) = M_v(\text{tag}, \text{why}, \text{src}) \in \{0, 1\}
\end{equation}
Only samples passing verification are included in the final training set $\mathbb{B} = \mathbb{B}^+ \cup \mathbb{B}^-$.

\subsection{Multi-Stage Model Training}

We adopt a two-stage training strategy to cultivate the Data-centric Structuring Agent (DSA): structure-aware supervised fine-tuning (SFT) establishes format compliance and basic discrimination capabilities, followed by reinforcement learning with multi-dimensional rewards to further optimize decision quality.

\subsubsection{Structure-Aware Supervised Fine-Tuning}

In the SFT stage, we fine-tune the base model using the constructed dataset $\mathbb{B}$ to master the generation of DSA-Lite Schema.

Considering that auditing decisions (Accept/Reject) are the core outputs of DSA, we apply loss weighting to the tag tokens:
\begin{equation}
\mathcal{L}_{\text{SFT}} = -\sum_{t=1}^{T} w_t \log p_\theta(y_t | y_{<t}, x)
\end{equation}
where $p_\theta(y_t | y_{<t}, x)$ represents the probability of predicting the $t$-th token given the input $x$ and the previously generated prefix $y_{<t}$ under model parameters $\theta$. The weight $w_t$ is defined as:
\begin{equation}
w_t = \begin{cases}
\alpha, & \text{if } y_t \in \{\text{Accept}, \text{Reject}, \text{tag}, \text{why}\} \\
1, & \text{otherwise}
\end{cases}
\end{equation}
We set $\alpha = 2.0$ to force the model to focus on learning classification decision boundaries.

\subsubsection{Group Policy Optimization with Multi-Dimensional Rewards}

In the reinforcement learning stage, we adopt the Group Relative Policy Optimization (GRPO) algorithm and design a multi-dimensional reward function to comprehensively guide model behavior.

For model output $y$ and golden label $y^*$, the total reward comprises five dimensions:

\textbf{(1) Format Reward $\mathcal{R}_{\text{format}}$:} Checks whether the output is valid parseable JSON that complies with the schema:
\begin{equation}
\mathcal{R}_{\text{format}} = \begin{cases}
1, & \text{if JSON is valid and schema-compliant} \\
0, & \text{otherwise}
\end{cases}
\end{equation}

\textbf{(2) Retrieval Reward $\mathcal{R}_{\text{retrieval}}$:} Evaluates the accuracy of evidence auditing decisions using F1 score to balance precision and recall:
\begin{equation}
\mathcal{R}_{\text{retrieval}} = \frac{2 \cdot P \cdot R}{P + R}
\end{equation}
where $P = \frac{|\mathcal{S}_{\text{pred}}^+ \cap \mathcal{S}_{\text{gold}}^+|}{|\mathcal{S}_{\text{pred}}^+|}$, $R = \frac{|\mathcal{S}_{\text{pred}}^+ \cap \mathcal{S}_{\text{gold}}^+|}{|\mathcal{S}_{\text{gold}}^+|}$, and $\mathcal{S}^+$ denotes the set of evidence marked as ACCEPT.

\textbf{(3) Consistency Penalty $\mathcal{R}_{\text{consistency}}$:} Detects logical self-consistency and severely penalizes contradictory behavior:
\begin{equation}
\mathcal{R}_{\text{consistency}} = \begin{cases}
-\beta, & \text{if } \exists v: v \in \mathcal{S}_{\text{pred}}^- \land v \in \mathcal{V}_{\text{logic}} \\
0, & \text{otherwise}
\end{cases}
\end{equation}
where $\mathcal{V}_{\text{logic}}$ is the set of values referenced in the computation logic, and $\beta$ is set to a large value (e.g., 5.0) to heavily penalize logical contradictions.

\textbf{(4) Executability Reward $\mathcal{R}_{\text{exec}}$:} Verifies whether the generated computation logic can be correctly parsed and executed by the symbolic executor:
\begin{equation}
\mathcal{R}_{\text{exec}} = \begin{cases}
1, & \text{if } \mathcal{E}(\mathcal{L}) \text{ executes without error} \\
0, & \text{otherwise}
\end{cases}
\end{equation}

\textbf{(5) Accuracy Reward $\mathcal{R}_{\text{accuracy}}$:} Compares the execution result with the ground truth answer:
\begin{equation}
\mathcal{R}_{\text{accuracy}} = \begin{cases}
1, & \text{if } |a_{\text{pred}} - a_{\text{gold}}| < \epsilon \\
0, & \text{otherwise}
\end{cases}
\end{equation}

Combining these above values, we get the final reward:
\begin{equation}
\mathcal{R}_{\text{total}} = \mathcal{R}_{\text{format}} + \mathcal{R}_{\text{retrieval}} + \mathcal{R}_{\text{exec}} + \mathcal{R}_{\text{accuracy}} + \mathcal{R}_{\text{consistency}}
\end{equation}

For each input sample $x \in \mathbb{B}$, we sample $G$ candidate outputs $\{y_1, y_2, \cdots, y_G\}$ from the current policy $\pi_\theta$. For each output $y_i$, we compute its reward $\mathcal{R}(y_i)$ and obtain the advantage function through group normalization: $A_i = \frac{\mathcal{R}(y_i) - \text{mean}(\{\mathcal{R}(y_j)\}_{j=1}^G)}{\text{std}(\{\mathcal{R}(y_j)\}_{j=1}^G)}$. The optimization objective is defined as:
\begin{equation}
\begin{aligned}
J_{\text{GRPO}}(\theta) = &\mathbb{E}_{x \sim \mathbb{B}} \mathbb{E}_{y \sim \pi_\theta} \Big[ \min\Big(\frac{\pi_\theta(y|x)}{\pi_{\text{old}}(y|x)} A_i, \\
&\text{clip}\Big(\frac{\pi_\theta(y|x)}{\pi_{\text{old}}(y|x)}, 1-\epsilon, 1+\epsilon\Big) A_i \Big) \\ &- \beta \text{KL}(\pi_\theta \| \pi_{\text{ref}}) \Big]
\end{aligned}
\end{equation}
where $\pi_{\text{old}}$ is the policy from the previous iteration, $\epsilon$ is the clipping threshold that constrains the magnitude of policy updates, $\beta$ is the KL divergence penalty coefficient, and $\pi_{\text{ref}}$ is the reference policy (typically the initial policy after SFT) that prevents the policy from deviating too far.

\subsection{Compile-and-Execute Inference}

After training, DSA serves as a deterministic compiler during inference, transforming user queries and retrieved documents into executable structured programs, which are subsequently symbolically executed to produce verifiable answers.

\subsubsection{DSA Representation}

The intermediate representation $\mathcal{I}$ generated by DSA comprises three core components. The first is task decomposition, which decomposes complex queries into atomic sub-tasks:
\begin{equation}
\mathcal{T} = \{T_i = (\text{target}_i, \text{constraint}_i)\}_{i=1}^{m}
\end{equation}
The second is the shared evidence pool, where each piece of evidence carries multi-head decision routing, i.e., discrimination results for the same evidence under different task perspectives:
\begin{equation}
\mathcal{P} = \{e_j = (v_j, \text{src}_j, \{(T_i, \text{tag}_{ij}, \text{why}_{ij})\})\}_{j=1}^{n}
\end{equation}
The third is symbolic computation logic, which references operators from a pre-registered operator library:
\begin{equation}
\mathcal{L} = \text{op}(T_1, T_2, \cdots)
\end{equation}

\subsubsection{Deterministic Symbolic Execution}

DSA executes the compiled DSA-Lite as a static computational graph through three steps. First, it performs multi-head routing aggregation by traversing the evidence pool and assigning evidence to corresponding task buckets $\mathcal{B}_i$ based on decision tags, while generating an audit log. Second, it conducts symbol binding and conflict detection through consistency verification on each task bucket:
\begin{equation}
\text{Bind}(T_i) = \begin{cases}
v, & \text{if } |\mathcal{B}_i| = 1 \\
\text{CONFLICT}, & \text{if } |\{v : v \in \mathcal{B}_i\}| > 1 \\
\text{MISSING}, & \text{if } |\mathcal{B}_i| = 0
\end{cases}
\end{equation}
Third, it invokes the pre-registered operator library within a safe sandbox for computation, completely eliminating code injection risks:
\begin{equation}
a = \text{OPERATOR\_REGISTRY.execute}(\mathcal{L}, \{\text{Bind}(T_i)\})
\end{equation}

The entire process generates a complete audit trail, recording the acceptance reason or rejection rationale for each piece of evidence, achieving fine-grained traceability.

For example: ``Accepted evidence: 59.1 ([Exact\_Match]), 98.0 ([Exact\_Match]) | Rejected evidence: 8.2 ([Concept\_Shift]) | Computation logic: divide(59.1, 98.0) | Result: 0.60306.''

This level of explainability is difficult to achieve with traditional Chain-of-Thought methods.

\section{Offline experiments}

To validate the effectiveness of DCRC (Data-Centric Reasoning Compiler), we conduct comprehensive offline experiments on financial numerical reasoning benchmarks. We seek to answer the following research questions:
\begin{itemize}[leftmargin=*]
\item \textbf{RQ1}: How does DCRC perform compared to existing SOTA methods for financial numerical reasoning?

\item \textbf{RQ2}: How does the DCRC compiler paradigm compare to the traditional CoT generation paradigm?

\item \textbf{RQ3}: How effective is the system's auditability and intermediate process quality?

\item \textbf{RQ4}: What are the contributions of key components in DCRC?

\item \textbf{RQ5}: How do key training hyperparameters affect model performance?

\end{itemize}

\subsection{Experimental settings}

\subsubsection{Datasets}
We adopt two financial numerical reasoning benchmarks. \textbf{FinQA}~\cite{finqa} contains 8,281 QA pairs (train 6,251/dev 883/test 1,147), involving table-text hybrid reasoning (62.43\% table-only, 23.42\% text-only, 14.15\% hybrid). \textbf{ConvFinQA}~\cite{convfinqa} contains 3,892 dialogues with 14,115 questions (train 3,037/dev 421/test 434), requiring multi-turn reasoning. We use the training sets from both datasets for model training and evaluate on FinQA's dev and test sets along with ConvFinQA's dev set (ConvFinQA test answers are private).

\subsubsection{Metrics}
We adopt \textbf{Execution Accuracy} as the primary metric, measuring the match rate between predicted and ground-truth results (with 1\% tolerance). To evaluate DSA's auditability, we propose three metrics: \textbf{Noise Filtering Precision} measures the proportion of true distractors among rejected evidence; \textbf{Evidence Retrieval Precision} measures the proportion of correct items among accepted evidence; \textbf{Audit Logic Accuracy} measures whether rejection reasons conform to the predefined error taxonomy.

\subsubsection{Baselines}
We compare with methods spanning three categories. \textbf{Prompting-based methods} include Few-shot CoT, which applies traditional chain-of-thought reasoning to elicit step-by-step solutions. \textbf{Program-based methods} include FinQANet~\cite{finqa}, which generates domain-specific programs from financial documents; CBR-Ren~\cite{cbr}, which leverages case-based reasoning for program retrieval and adaptation; and PoT-GPT-4~\cite{pot-gpt-4}, which employs Program-of-Thought prompting with GPT-4 as the backbone. \textbf{Retriever-generator methods} include ArgRecog, which focuses on argument recognition for numerical reasoning; ENCORE~\cite{encore}, which derives reliable reasoning through answer formula decomposition; APOLLO~\cite{apollo}, which enhances retrieval via number-aware negative sampling; and FINDER~\cite{finder}, which combines generative retrieval with dynamic in-context example selection. For fair comparison, FINDER uses Mistral-7B Instruct v0.2 (7B parameters) as its generator, matching the parameter scale of our DSA-7B model.

\subsubsection{Implementation Details}
We train DSA based on Qwen2.5 series (1.5B/3B/7B). We analyze the ACCEPT/REJECT sample distribution in the above training sets and synthesize the final training data with a 1:2 positive-negative sample ratio for data augmentation. Training consists of two stages: \textbf{Stage 1} employs structure-aware SFT with $\lambda=2.0$ weight scaling on ACCEPT/REJECT tokens, training for 3 epochs with learning rate 5e-6 and batch size 64; \textbf{Stage 2} applies GRPO reinforcement learning, sampling $G=16$ outputs per input and optimizing with multi-dimensional rewards (format, retrieval F1, consistency, execution accuracy) for 3 epochs with learning rate 1e-6.

\begin{table}[!h]
    \centering
    \caption{Comparison with SOTA methods. Execution Accuracy (\%) is reported.}
    \label{tab:sota}
    \scalebox{0.9}{
    \begin{tabular}{l|c c c}
    \toprule
    \textbf{Method} & \textbf{FinQA-Dev} & \textbf{FinQA-Test} & \textbf{ConvFinQA-Dev} \\
    \midrule
    \multicolumn{4}{l}{\textit{Base Models (Zero-shot)}} \\
    \midrule
    Qwen2.5-1.5B-Ins & 35.10 & 34.22 & 30.15 \\
    Qwen2.5-3B-Ins & 48.30 & 47.50 & 44.80 \\
    Qwen2.5-7B-Ins & 61.50 & 60.80 & 59.20 \\
    \midrule
    \multicolumn{4}{l}{\textit{Existing SOTA Methods}} \\
    \midrule
    FinQANet & 61.22 & 61.24 & 68.32 \\
    ArgRecog & 67.50 & 64.86 & 73.94 \\
    CBR-Ren & 68.40 & 67.81 & 72.61 \\
    PoT-GPT-4 & 71.05 & 69.38 & 74.77 \\
    ENCORE & 71.60 & 69.40 & 76.00 \\
    APOLLO & 72.91 & 71.07 & 78.76 \\
    FINDER & 77.13 & 75.32 & 81.95 \\
    \midrule
    \multicolumn{4}{l}{\textit{DCRC (Ours)}} \\
    \midrule
    \rowcolor{gray!15} DSA-1.5B & 73.05 & 73.50 & 71.00 \\
    \rowcolor{gray!15} DSA-3B & 79.45 & 79.12 & 78.50 \\
    \rowcolor{gray!15} DSA-7B & \textbf{84.71} & \textbf{84.66} & \textbf{85.67} \\
    \bottomrule
    \end{tabular}
    }
\end{table}

\subsection{RQ1: Comparison with SOTA Methods}

To evaluate the performance gap between DCRC and existing methods, we compare DSA with various baselines, including traditional chain-of-thought methods (Few-shot CoT), program generation approaches (FinQANet), and recent SOTA methods (ENCORE, APOLLO, FINDER). The results are presented in Table~\ref{tab:sota}.

Based on the results, we make the following observations. First, even our smallest 1.5B model (73.50\%) surpasses ENCORE (69.40\%) and APOLLO (71.07\%), demonstrating the efficiency advantage of the compiler paradigm through decoupling reasoning from computation. Second, our 7B model achieves 84.66\% on FinQA-Test, improving upon the previous best FINDER (75.32\%) by 9.34 percentage points, and reaches 85.67\% on ConvFinQA with significant lead, validating the effectiveness of the DCRC architecture. Finally, DCRC achieves best performance on both FinQA and ConvFinQA, indicating good generalization ability across different financial QA scenarios.

\subsection{RQ2: Compiler Paradigm vs. CoT Generation Paradigm}

To verify the effectiveness of the structured compilation paradigm, we compare CoT (chain-of-thought) and DCRC (our method) across different model scales. CoT relies on implicit reasoning through free-text generation, while DCRC confines LLM uncertainty to the compilation phase (evidence filtering and structuring) while keeping the execution phase (numerical computation) deterministic. The results are presented in Table~\ref{tab:paradigm}.

Based on the results, we draw the following key findings. First, even without any training, the data-centric compilation paradigm of DCRC demonstrates superiority over the CoT paradigm---on the 1.5B base model, DCRC achieves 35.10\% while CoT only reaches 32.45\%, indicating that structured output itself guides more precise information extraction. Second, after SFT+RL training, models trained with DCRC significantly outperform those trained with CoT---at 7B scale, DCRC achieves 84.66\% accuracy, surpassing CoT's 77.50\% by 7.16 percentage points, validating that training models to generate verifiable structured programs is a superior paradigm. Furthermore, DCRC's advantage persists across all model scales, demonstrating good scalability.

\begin{table*}[!h]
    \centering
    \caption{Comparison of reasoning paradigms across model scales. Execution Accuracy (\%) is reported.}
    \label{tab:paradigm}
    \scalebox{0.9}{
    \begin{NiceTabular}{c|c|c|c c c}
    \toprule
    \textbf{Size} & \textbf{Stage} & \textbf{Method} & \textbf{FinQA-Dev} & \textbf{FinQA-Test} & \textbf{ConvFinQA-Dev} \\
    \midrule
    \Block{4-1}{Qwen2.5-1.5B-Instruct} 
        & \Block{2-1}{Base Model} 
            & Few-shot CoT & 32.45 & 31.80 & 28.50 \\
        &   & \rowcolor{gray!15} DCRC & \textbf{35.10} & \textbf{34.22} & \textbf{30.15} \\
        \cmidrule{2-6}
        & \Block{2-1}{SFT+RL} 
            & CoT & 65.20 & 64.85 & 62.10 \\
        &   & \rowcolor{gray!15} DCRC & \textbf{73.05} & \textbf{73.50} & \textbf{71.00} \\
    \midrule
    \Block{4-1}{Qwen2.5-3B-Instruct} 
        & \Block{2-1}{Base Model} 
            & Few-shot CoT & 45.60 & 44.90 & 41.20 \\
        &   & \rowcolor{gray!15} DCRC & \textbf{48.30} & \textbf{47.50} & \textbf{44.80} \\
        \cmidrule{2-6}
        & \Block{2-1}{SFT+RL} 
            & CoT & 72.50 & 71.80 & 70.50 \\
        &   & \rowcolor{gray!15} DCRC & \textbf{79.45} & \textbf{79.12} & \textbf{78.50} \\
    \midrule
    \Block{4-1}{Qwen2.5-7B-Instruct} 
        & \Block{2-1}{Base Model} 
            & Few-shot CoT & 58.40 & 57.20 & 55.60 \\
        &   & \rowcolor{gray!15} DCRC & \textbf{61.50} & \textbf{60.80} & \textbf{59.20} \\
        \cmidrule{2-6}
        & \Block{2-1}{SFT+RL} 
            & CoT & 78.20 & 77.50 & 78.10 \\
        &   & \rowcolor{gray!15} DCRC & \textbf{84.71} & \textbf{84.66} & \textbf{85.67} \\
    \bottomrule
    \end{NiceTabular}
}
\end{table*}

\subsection{RQ3: Auditability Evaluation}

\begin{table}[!h]
    \centering
    \caption{Auditability and intermediate process quality evaluation on FinQA-Test (\%).}
    \label{tab:audit}
    \scalebox{0.9}{
    \begin{tabular}{c|c c c}
    \toprule
    \textbf{Model} & \makecell[c]{\textbf{Noise Filtering}\\\textbf{Precision}} & \makecell[c]{\textbf{Evidence Retrieval}\\\textbf{Precision}} & \makecell[c]{\textbf{Audit Logic}\\\textbf{Accuracy}} \\
    \midrule
    DSA-1.5B & 86.4 & 88.2 & 92.5 \\
    DSA-3B & 90.1 & 91.5 & 97.8 \\
    \rowcolor{gray!15} DSA-7B & \textbf{97.2} & \textbf{96.4} & \textbf{99.6} \\
    \bottomrule
    \end{tabular}
    }
\end{table}

Auditability is a core design goal of DCRC. Unlike traditional CoT methods that only demonstrate ``how to get it right'', DCRC explicitly records ``why incorrect options were rejected'', providing complete audit trails for financial applications. Table~\ref{tab:audit} validates DCRC's auditability from three dimensions.

The 7B model achieves 97.2\% Noise Filtering Precision, accurately identifying and rejecting distractors—highly similar terms like ``net profit attributable to parent'' vs ``net profit'' are main error sources in traditional methods, which DCRC effectively addresses through explicit adversarial discrimination. Evidence Retrieval Precision reaches 96.4\%, confirming that DSA-accepted evidence is almost always correct, preventing hallucinated data from entering computations. Audit Logic Accuracy achieves 99.6\%, demonstrating that rejection reasons almost perfectly align with the predefined error taxonomy (e.g., \texttt{[Time\_Mismatch]}, \texttt{[Entity\_Mismatch]}) rather than vague free-text explanations. These results demonstrate that DCRC not only produces correct answers but also completely explains the decision process—this ``Negative Reasoning'' capability is difficult for traditional CoT methods to provide.

\subsection{RQ4: Ablation Study}

\begin{table}[!h]
    \centering
    \caption{Ablation study on FinQA-Test. Execution Accuracy (\%) is reported. Numbers in parentheses indicate absolute drops.}
    \label{tab:ablation}
    \scalebox{0.9}{
    \begin{tabular}{c|c c c c c}
    \toprule
    \textbf{Model} & \textbf{Full} & \makecell[c]{\textbf{w/o Adv.}\\\textbf{Audit}} & \makecell[c]{\textbf{w/o Neg.}\\\textbf{Injection}} & \makecell[c]{\textbf{w/o Multi-}\\\textbf{Dim Reward}} & \makecell[c]{\textbf{w/o Struct.}\\\textbf{SFT}} \\
    \midrule
    1.5B & 73.50 & 64.10 (-9.4) & 69.10 (-4.4) & 70.50 (-3.0) & 68.20 (-5.3) \\
    3B & 79.12 & 70.20 (-8.9) & 75.20 (-3.9) & 76.80 (-2.3) & 74.50 (-4.6) \\
    7B & 84.66 & 77.50 (-7.1) & 81.30 (-3.3) & 82.90 (-1.7) & 80.15 (-4.5) \\
    \bottomrule
    \end{tabular}
    }
\end{table}

We systematically remove key components of DCRC to evaluate their contributions. Specifically, we ablate adversarial auditing (ICCD), structured SFT, negative injection, and multi-dimensional rewards respectively. The results are shown in Table~\ref{tab:ablation}.

Adversarial auditing (ICCD) contributes most significantly, with removal causing the largest drop (7B: -7.1\%), proving that In-Context Contrastive Denoising is the core mechanism—without explicit adversarial discrimination training, the model is easily confused by similar distractors. Structured SFT provides a necessary foundation for format and classification logic before reinforcement learning (7B: -4.5\%). Negative injection enhances generalization (7B: -3.3\%), as hard negatives force fine-grained discrimination rather than simple pattern matching. Although multi-dimensional rewards cause relatively smaller drop (7B: -1.7\%), further analysis reveals their crucial role in maintaining internal logical coherence—cases of ``REJECT-marked evidence appearing in final computations'' increase significantly without them. These results demonstrate that DCRC components work synergistically to ensure superior performance in both accuracy and logical consistency.

\subsection{RQ5: Hyperparameter Sensitivity Analysis}

\begin{figure}[!t]
    \centering
    \includegraphics[width=\linewidth]{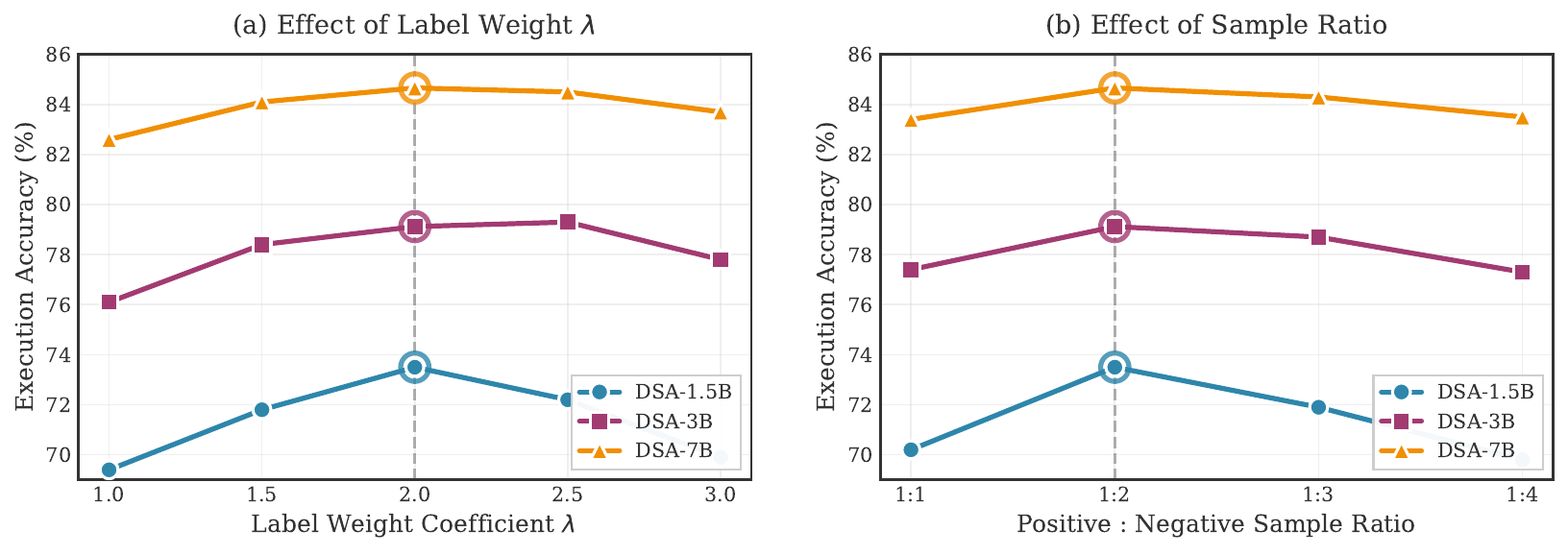}
    \caption{Hyperparameter sensitivity analysis on FinQA-Test. 
    (a) Effect of label weight coefficient $\lambda$; (b) Effect of positive-negative sample ratio. Circles highlight optimal configurations.
    }
    \label{fig:hyperparameter}
\end{figure}

We analyze two key hyperparameters: the label weight coefficient $\lambda$ in structured SFT and the positive-negative sample ratio in adversarial data synthesis. Results are shown in Figure~\ref{fig:hyperparameter}.

Both hyperparameters exhibit inverted-U curves, with smaller models showing higher sensitivity. For $\lambda$, the 1.5B model shows sharp drops at extreme values (69.40\% at $\lambda$=1.0 vs 73.50\% at $\lambda$=2.0), while larger models remain relatively stable. For the sample ratio, excessive negatives (1:4) cause significant degradation especially for smaller models due to over-rejection of valid evidence. Based on comprehensive analysis, we adopt $\lambda=2.0$ and 1:2 ratio as default configurations, achieving optimal or near-optimal performance across all scales.

\section{Online Experiments}

To validate DCRC's effectiveness in production, we deploy it in Tencent Yuanbao's financial QA system for A/B testing.

\subsection{Online System}

\begin{figure}[!htbp]
    \centering
    \includegraphics[width=0.7\linewidth,height=0.35\textheight]{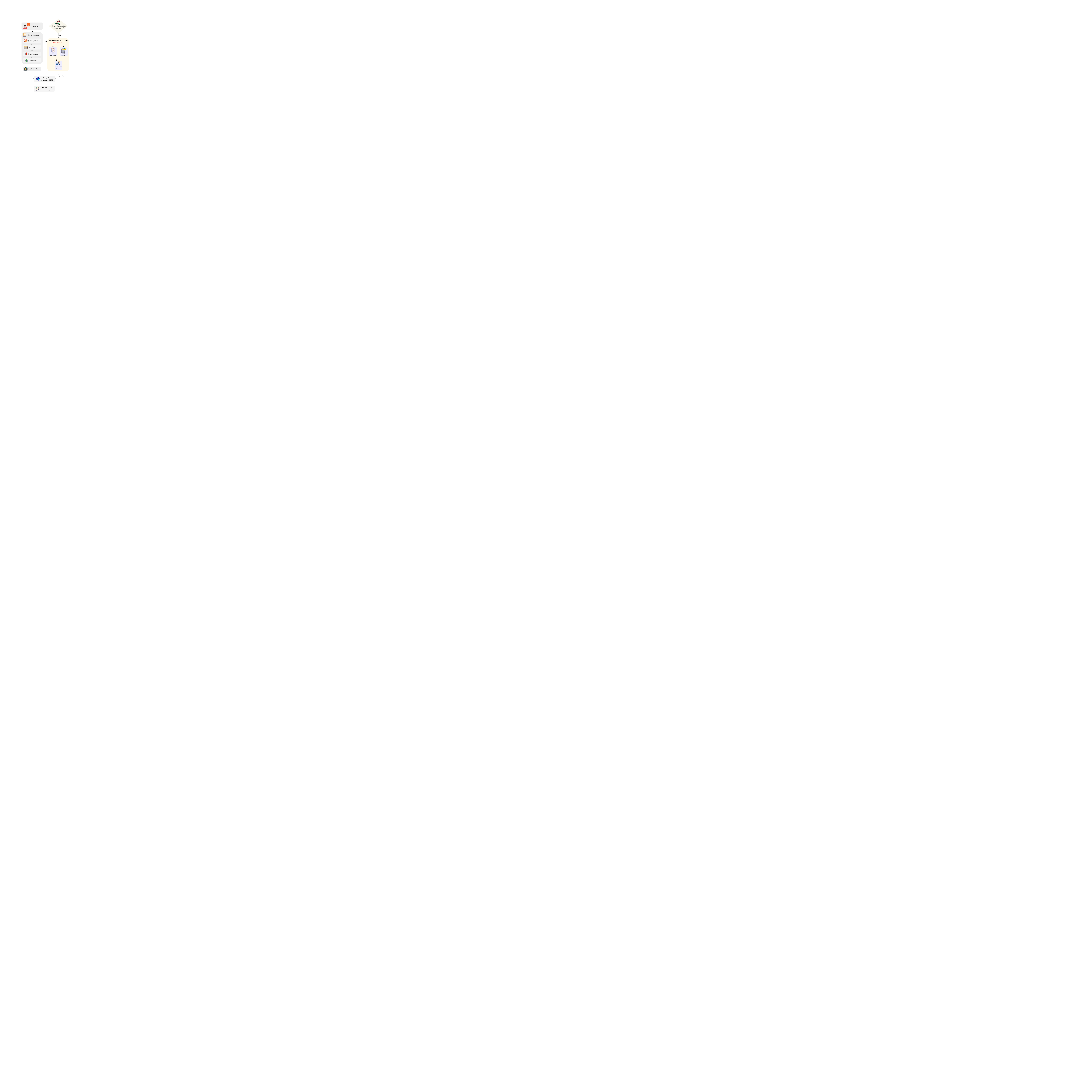}
    \caption{Architecture of the online financial QA system. Blue indicates the main pipeline; orange indicates the DSA enhancement branch (our contribution).}
    \label{fig:online_system}
\end{figure}

As shown in Figure~\ref{fig:online_system}, the system consists of a main pipeline and an enhancement branch. The main pipeline retrieves top-$k$ documents through the retrieval module (query expansion, tool calling, coarse/fine ranking) and generates answers via a large-scale LLM. The enhancement branch activates the DSA module for queries identified as financial numerical reasoning, processing retrieval results into structured evidence with audit annotations and program execution results to enhance generator input. Each DSA module is deployed on 2$\times$H20 GPUs (vLLM), with average output of 182 tokens and latency of 0.92s (range: 0.37s-1.24s).

\subsection{Online Results}

\begin{table}[!t]
    \centering
    \caption{Online A/B results (7 days, 1,000 samples per group).}
    \label{tab:online}
    \begin{tabular}{l|cc|c}
    \toprule
    \textbf{Metric} & \textbf{Baseline} & \textbf{+DSA} & \textbf{$\Delta$} \\
    \midrule
    Hallucination Rate $\downarrow$ & 12.44\% & 6.82\% & -5.62\% \\
    Answer Accuracy $\uparrow$ & 85.26\% & 91.48\% & +6.22\% \\
    \midrule
    Real-time Metric Acc. $\uparrow$ & 87.79\% & 95.69\% & +7.90\% \\
    Historical Metric Acc. $\uparrow$ & 86.73\% & 93.37\% & +6.64\% \\
    \midrule
    User Satisfaction $\uparrow$ & 72.00\% & 88.00\% & +16.00\% \\
    \bottomrule
    \end{tabular}
\end{table}

We conduct a 7-day A/B test with 9:1 traffic allocation for financial numerical reasoning queries, manually evaluating 2,000 fact claims (1,000 per group). As shown in Table~\ref{tab:online}, we report results across three dimensions: faithfulness, numerical accuracy, and user experience.

\textbf{Faithfulness.} DSA reduces hallucination rate from 12.44\% to 6.82\% (a relative reduction of 45.2\%) and improves overall answer accuracy from 85.26\% to 91.48\%. The accuracy gain stems from both reduced hallucination and DSA's explicit rejection of retrieval noise through structured auditing.

\textbf{Numerical Accuracy.} We further decompose numerical correctness into real-time metrics (e.g., current stock prices, latest quarterly earnings) and historical metrics (e.g., year-over-year growth rates, past financial indicators). DSA achieves substantial improvements on both: real-time metric accuracy increases from 87.79\% to 95.69\% (+7.90\%), while historical metric accuracy improves from 86.73\% to 93.37\% (+6.64\%). The gains on real-time metrics are particularly notable, as these queries are more susceptible to retrieval noise due to frequently updated data sources.

\textbf{User Satisfaction.} Human evaluators assess overall response quality considering correctness, completeness, and explainability. User satisfaction improves significantly from 72\% to 88\% (+16\%), indicating that DSA's structured audit logs and transparent reasoning enhance user trust beyond mere accuracy improvements.

Trading 0.92s additional latency for these significant quality improvements represents a reasonable trade-off in high-stakes financial scenarios where accuracy and trustworthiness are paramount. 
\section{Conclusion}

In this paper, we present the Data-centric Reasoning Compiler (DCRC), a novel framework that addresses numerical hallucinations in online financial QA through a data-centric paradigm. DCRC comprises adversarial data construction, multi-stage agent training, and compile-and-execute inference, effectively tackling noise sensitivity, calculation fragility, and auditability crisis.
We hope our work provides practitioners with practical experience in deploying trustworthy LLMs for high-stakes financial applications, and that the data-centric approach offers a new perspective beyond traditional model-centric optimizations. In the future, we plan to extend our framework to support more complex reasoning scenarios, such as multi-hop cross-document reasoning and real-time streaming financial data analysis.

\section{Acknowledgement}
We thank the support of the Shenzhen Technology University School-level
(No.20251061020002)
\bibliographystyle{ACM-Reference-Format}
\bibliography{reference}
\clearpage
\appendix
\onecolumn
\section{Prompt used in Method}

\begin{tcolorbox}[
    colback=blue!5!white,
    colframe=blue!75!black,
    title=\textbf{Prompt: DSA (Data-centric Structuring Agent)},
    fonttitle=\bfseries,
    breakable,
    enhanced,
    width=\textwidth,
    left=2mm,
    right=2mm,
    top=2mm,
    bottom=2mm,
    label=fig:dsa-prompt
]
\small\ttfamily
\textbf{\#\# Role:} \\
You are a professional financial numerical reasoning analyst. Your task is to analyze financial documents and tables to answer users' financial numerical calculation questions. \\

\textbf{\#\# Task Description:} \\
Given a financial document (containing text and tables) and a question, you need to: \\
1. \textbf{Understand the question}: Identify the target metric and constraints required for calculation \\
2. \textbf{Extract evidence}: Extract all potentially relevant/confusable numerical evidence from the document \\
3. \textbf{Discriminate and filter}: For each piece of evidence, decide whether to accept it with structured reasons  \\
4. \textbf{Construct calculation}: Determine computation logic and output the final answer \\

\textbf{\#\# Output Format:} \\
Please strictly follow the JSON format below:
\begin{verbatim}
{
  "tasks": {
    "T1": { "target": "target metric name", "constraint": "time/condition" },
    "T2": { "target": "target metric name", "constraint": "time/condition" }
  },
  "pool": [
    {
      "id": "E1",
      "val": "value (keep original format)",
      "src": "source text snippet",
      "decisions": {
        "T1": { "tag": "ACCEPT", "why": "[Exact_Match]" },
        "T2": { "tag": "REJECT", "why": "[Time_Mismatch]" }
      }
    }
  ],
  "logic": "calculation expression"
}
\end{verbatim}
\vspace{\baselineskip}

\textbf{\#\# Decision Tag Descriptions:} \\
\textbf{ACCEPT Reason Codes:} \\
\hspace*{2em}- \texttt{[Exact\_Match]} - Exactly matches target metric and all constraints \\
\hspace*{2em}- \texttt{[Derived\_Match]} - Can be derived through simple transformation \\

\textbf{REJECT Reason Codes:} \\
\hspace*{2em}- \texttt{[Time\_Mismatch]} - Time/year does not match \\
\hspace*{2em}- \texttt{[Entity\_Mismatch]} - Entity/company name does not match \\
\hspace*{2em}- \texttt{[Concept\_Shift]} - Different metric concept (e.g., operating profit vs. net profit) \\
\hspace*{2em}- \texttt{[Unit\_Error]} - Unit mismatch \\
\hspace*{2em}- \texttt{[Irrelevant]} - Value irrelevant to the question \\

\textbf{\#\# Supported Calculation Operations:} \\
\hspace*{2em}- \texttt{add(a, b)} - Addition \\
\hspace*{2em}- \texttt{subtract(a, b)} - Subtraction (a - b) \\
\hspace*{2em}- \texttt{multiply(a, b)} - Multiplication \\
\hspace*{2em}- \texttt{divide(a, b)} - Division (a / b) \\
\hspace*{2em}- \texttt{exp(a, b)} - Exponentiation (a \^{} b) \\
\hspace*{2em}- \texttt{greater(a, b)} - Comparison (returns "yes" or "no") \\

Use \texttt{\#N} to reference the result of step N (starting from 0), e.g.: \\
\hspace*{2em}- \texttt{subtract(100, 80), divide(\#0, 80)} means first compute 100-80=20, then 20/80=0.25 \\


\end{tcolorbox}

\section{Illustrative Example: DSA Evidence Auditing}

\begin{tcolorbox}[
    colback=purple!5!white,
    colframe=purple!75!black,
    title=\textbf{Example: DSA Output for Financial Numerical Reasoning},
    fonttitle=\bfseries,
    breakable,
    enhanced,
    width=\textwidth,
    left=2mm,
    right=2mm,
    top=2mm,
    bottom=2mm,
    label=fig:dsa-example
]
\small\ttfamily
\textbf{\#\# Question:} \\
What was the difference in percentage return for PMI common stock compared to the S\&P 500 index for the five years ended 2018? \\

\textbf{\#\# Gold Answer:} -0.538 \\

\textbf{\#\# DSA Output:}
\begin{verbatim}
{
  "tasks": {
    "T1": { "target": "pmi", "constraint": "december 31 2018" },
    "T2": { "target": "s&p 500 index", "constraint": "december 31 2018" }
  },
  "pool": [
    { "id": "E1", "val": "96.50", 
      "src": "december 31 2018 | $96.50 | $127.70 | $150.30",
      "decisions": { 
        "T1": { "tag": "ACCEPT", "why": "[Exact_Match]" },
        "T2": { "tag": "REJECT", "why": "[Concept_Shift]" } } },
    { "id": "E2", "val": "150.30", 
      "src": "december 31 2018 | $96.50 | $127.70 | $150.30",
      "decisions": { 
        "T1": { "tag": "REJECT", "why": "[Concept_Shift]" },
        "T2": { "tag": "ACCEPT", "why": "[Exact_Match]" } } },
    { "id": "E3", "val": "100.00", 
      "src": "december 31 2013 | $100.00 | $100.00 | $100.00",
      "decisions": { "ALL": { "tag": "REJECT", "why": "[Time_Mismatch]" } } },
    { "id": "E4", "val": "127.70", 
      "src": "december 31 2018 | $96.50 | $127.70 | $150.30",
      "decisions": { "ALL": { "tag": "REJECT", "why": "[Concept_Shift]" } } },
    { "id": "E5", "val": "144.50", 
      "src": "december 31 2017 | $144.50 | $140.50 | $157.20",
      "decisions": { "ALL": { "tag": "REJECT", "why": "[Time_Mismatch]" } } }
  ],
  "logic": "subtract(96.50, const_100), divide(#0, const_100), 
           subtract(150.30, const_100), divide(#2, const_100), 
           subtract(#1, #3)"}
\end{verbatim}
\vspace{\baselineskip}

\textbf{\#\# Audit Log:} \\
Accepted evidence: 96.50 ([Exact\_Match]), 150.30 ([Exact\_Match]) | Rejected evidence: 96.50 ([Concept\_Shift]), 150.30 ([Concept\_Shift]), 100.00 ([Time\_Mismatch]), 127.70 ([Concept\_Shift]), 144.50 ([Time\_Mismatch]) | Computation logic: subtract(96.50, const\_100), divide(\#0, const\_100), subtract(150.30, const\_100), divide(\#2, const\_100), subtract(\#1, \#3) | Result: -0.538

\end{tcolorbox}

\clearpage

\section{Illustrative Example: DSA Program Execution}
\begin{tcolorbox}[
    colback=orange!5!white,
    colframe=orange!75!black,
    title=\textbf{Operator Registry and Deterministic Execution Process},
    fonttitle=\bfseries,
    breakable,
    enhanced,
    width=\textwidth,
    left=2mm,
    right=2mm,
    top=2mm,
    bottom=2mm,
    label=fig:operator-registry
]
\small\ttfamily
\textbf{\#\# Registered Operator Library:}
\begin{verbatim}
OPERATOR_REGISTRY = {
    "add":       lambda a, b: a + b,
    "subtract":  lambda a, b: a - b,
    "multiply":  lambda a, b: a * b,
    "divide":    lambda a, b: a / b,
    "exp":       lambda a, b: a ** b,
    "greater":   lambda a, b: "yes" if a > b else "no",
    "table_max": lambda row: max(row),
    "table_min": lambda row: min(row),
    "table_sum": lambda row: sum(row),
    "table_avg": lambda row: sum(row) / len(row)
}
\end{verbatim}
\vspace{\baselineskip}

\textbf{\#\# Execution Example:} \\
\textbf{Input Program:} \texttt{subtract(96.50, const\_100), divide(\#0, const\_100), ...} \\

\textbf{Step-by-Step Execution:}
\begin{verbatim}
# Step 0: subtract(96.50, const_100)
>>> arg1, arg2 = 96.50, 100.0
>>> res[0] = OPERATOR_REGISTRY["subtract"](arg1, arg2)
>>> res[0] = -3.50

# Step 1: divide(#0, const_100)
>>> arg1, arg2 = res[0], 100.0  # #0 references Step 0
>>> res[1] = OPERATOR_REGISTRY["divide"](arg1, arg2)
>>> res[1] = -0.035

# Step 2: subtract(150.30, const_100)
>>> arg1, arg2 = 150.30, 100.0
>>> res[2] = OPERATOR_REGISTRY["subtract"](arg1, arg2)
>>> res[2] = 50.30

# Step 3: divide(#2, const_100)
>>> arg1, arg2 = res[2], 100.0
>>> res[3] = OPERATOR_REGISTRY["divide"](arg1, arg2)
>>> res[3] = 0.503

# Step 4: subtract(#1, #3)
>>> arg1, arg2 = res[1], res[3]
>>> res[4] = OPERATOR_REGISTRY["subtract"](arg1, arg2)
>>> res[4] = -0.538

# Final Result: round(-0.538, 5) = -0.538
\end{verbatim}
\vspace{\baselineskip}

\textbf{\#\# Key Design Principles:} \\
\hspace*{2em}1. \textbf{Sandbox Execution}: All operations invoke pre-registered functions only, eliminating arbitrary code injection risks. \\
\hspace*{2em}2. \textbf{Reference Resolution}: \texttt{\#N} tokens are resolved to previous step results, enabling multi-step computation. \\
\hspace*{2em}3. \textbf{Deterministic Output}: Results are rounded to 5 decimal places for reproducibility.

\end{tcolorbox}


\end{document}